\documentclass[a4paper,11pt]{article}
\usepackage{pos}
\usepackage{tikz}
\usepackage{enumitem}
\usepackage{hyperref, url}

\usepackage{graphicx}
\usepackage{caption}
\usepackage{subcaption}
\usepackage{color}
\usepackage{titlesec}

\renewenvironment{thebibliography}[1]{
  \begin{oldthebibliography}{#1}
    \setlength{\itemsep}{0em}
    \setlength{\parskip}{0em}
    \setlength{\intextsep}{0em}
}
{
  \end{oldthebibliography}
}

\title{Are science exhibitions for everyone?\\Accessibility aspects of the CERN Science Gateway exhibitions}

\author*[a, b]{Tamara Caldas Cifuentes}
\author[a, c]{Jemma Harris}
\author[a]{Patricia Verheyden}

\affiliation[a]{Exhibitions team, CERN,\\
  Esplanade des Particules 1, 1217 Meyrin, Switzerland}

\affiliation[b]{Institute for Theoretical Physics, Goethe University,\\
Max-von-Laue-Straße 1, 60438 Frankfurt am Main, Germany}

\affiliation[c]{Science Communication Unit, Imperial College London,
Centre for Languages, Culture and Communication, London SW7 2AZ, UK}

\emailAdd{tamara.caldas.cifuentes@cern.ch}
\emailAdd{jemma.harris@cern.ch}
\emailAdd{patricia.verheyden@cern.ch}

\abstract{
  CERN’s new flagship education and outreach centre, \emph{Science Gateway}, opened its doors to the public in autumn 2023.
  Through a combination of immersive scenography with interactive exhibits and real scientific objects, its permanent exhibitions address the organisation’s particle physics research and how this knowledge applies to other scientific fields and the visitor’s everyday life.
  While in \emph{Discover CERN} visitors learn about the accelerator and detector technologies at CERN, \emph{Our Universe} covers the evolution of the Universe and the role that fundamental physics plays in it, as well as open questions in modern physics. Their underlying basic quantum mechanical principles can eventually be explored in \emph{Quantum World}. Firstly, we will give a short overview of the exhibitions’ highlights.
  Next, we delve into how various aspects of accessibility were incorporated into the exhibition development process to target the broadest audience possible and to inspire the next generation of scientists.
  We give an account on how this topic was approached and how inclusive exhibits and infrastructure were realised. 
  By combining long-term visitor statistics and preliminary results of a recent visitor survey, we obtain insights about the \emph{actual} audience and the visitors’ perceived accessibility of the exhibitions with respect to their educational backgrounds.
}

\definecolor{cern}{RGB}{0,83,161}

\ConferenceLogo{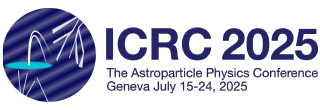}

\FullConference{39th International Cosmic Ray Conference (ICRC2025)\\
 15–24 July 2025\\
Geneva, Switzerland\\}

\begin{document}
\maketitle

\section{Introduction: A long history of research and outreach}
\label{sec:introduction}\noindent
CERN as an international institution is not only committed to excellence in fundamental physics research but also for its long history of diverse outreach activities. What started with the first on-site guided tours in the 1960s, developed into a broad spectrum of programmes. This includes the CERN \emph{Open Days}, the first permanent exhibition \emph{Microcosm}, the experimental lab workshop space \emph{S'Cool Lab} for teachers and students, and more.
The steadily increasing annual visitor numbers for the guided tours signalled a rising public interest in the institution and its research. Eventually, this led to the idea of the new flagship science outreach and education centre \emph{CERN Science Gateway}. Its new complementary educational programme consists of three permanent exhibitions (\emph{Discover CERN}, \emph{Our Universe}, and \emph{Quantum World}), the lab workshops, the science shows, and the guided visits.
Since its opening in October 2023, it has already welcomed more than 500 000 visitors from all over the world --- more precisely, 400 838 visitors in the reference period of June 2024 to May 2025. A statistical overview of this visitor cohort will be given in Sec.~\ref{sec:SG_statistical_overview}.\newline
Here, we present the permanent exhibitions (Sec.~\ref{sec:exhibitions_overview}) and comment on how and which aspects of accesibility have guided their creation process (Sec.~\ref{sec:accessibility_in_exhibition_development}). We conclude with statistical data on the visitor cohort described above and preliminary results from a recent internal visitor survey (Sec.~\ref{sec:stats_and_eval}). 
\section{Exhibitions overview: Exploring fundamental science in an interactive way}\label{sec:exhibitions_overview}
\begin{figure}[t]
    \begin{center}
      \includegraphics[width=\textwidth]{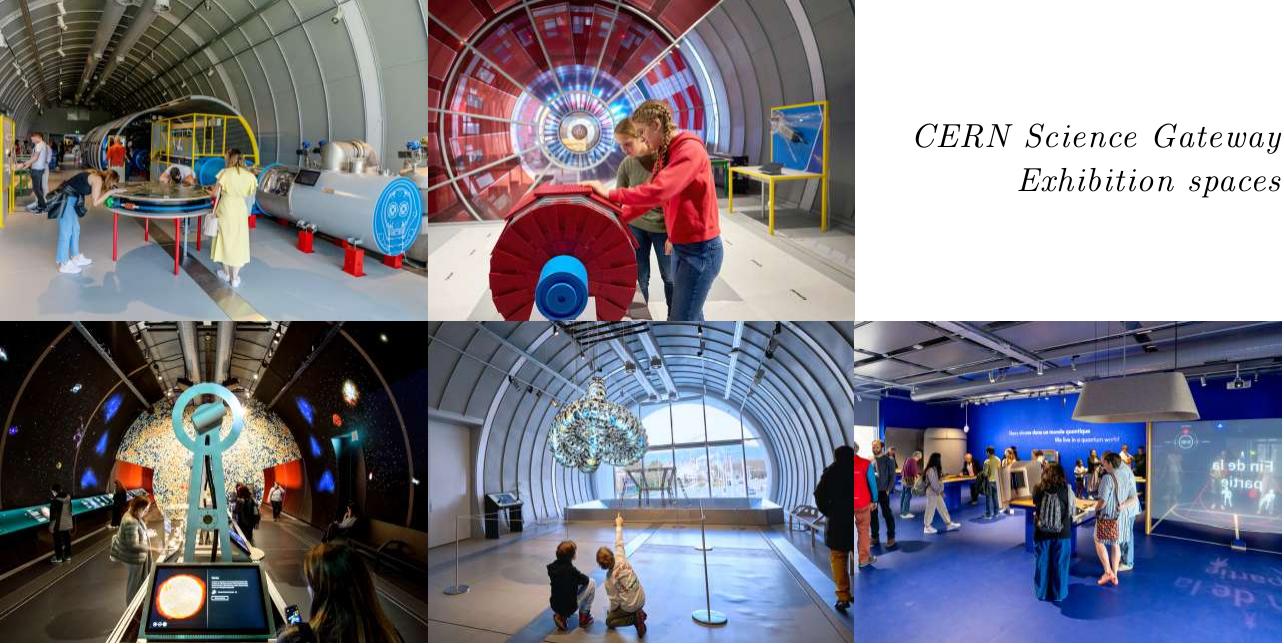}
      \caption{Overview of the five sub-exhibitions in CERN Science Gateway. \copyright CERN\\
      \emph{Upper left: Discover CERN: Accelerate}. \emph{Upper middle: Discover CERN: Collide}. \emph{Lower left: Our Universe: Back to the Big Bang}. \emph{Lower middle: Our Universe: Exploring the Unknown}. \emph{Lower right: Quantum World}.}
      \label{fig:collage_exh}
    \end{center}
  \end{figure}

\noindent
\emph{Science Gateway} houses three permanent exhibitions (see Fig.~\ref{fig:collage_exh}): \emph{Discover CERN}, \emph{Our Universe}, and \emph{Quantum World}. The first two are divided into two parts: \emph{Accelerate} and \emph{Collide}, and \emph{Back to the Big Bang} and \emph{Exploring the Unknown}, respectively.
All three stand out through their combination of immersive scenography with interactive exhibits (physical hands-on and multimedia type) as well as real scientific objects (parts of machinery or small-scale experiments). Visitors explore the exhibition spaces by themselves, while CERN volunteers provide guiding and further information on demand. 
The exhibitions demonstrate the relevance of CERN's research programme to advancing the field of particle physics as well as how this knowledge is applicable to other scientific disciplines and everyday life.
The exhibitions can be summarised as follows:
\vspace{-5pt}
\paragraph{Discover CERN} The \textbf{\emph{Accelerate}} exhibition is all about answering the question "What does one need to (understand to) build a machine like the LHC?". It explores the key technologies of CERN's accelerator infrastructure and the physical phenomena that power them. The predominant topics are electromagnetism (dipole/quadrupole magnets, RF cavities), superconductivity and civil engineering (tunnel construction). The highlight of the space is a life-size replica of the LHC tunnel that houses a real, functioning proton linear accelerator called \emph{ELISA} (\emph{Experimental Linac for Surface Analysis}), which has been designed for the exhibition as well as for research purposes. The \textbf{\emph{Collide}} part addresses the question "What does one need to (understand to) detect particles?", and features real detector parts alongside interactive exhibits on tracker systems and computational tasks such as data processing and analysis, for example. The exhibition space is sectioned by transparent walls with prints symbolising different slices through a particle detector; one of them uses real technical components to create a detector sculpture.
\vspace{-5pt}
\paragraph{Our Universe} While walking through the \textbf{\emph{Back to the Big Bang}} space, the visitors start in today's universe and follow its evolution backwards in time up until shortly before the Big Bang, guided by the question "Where do the macroscopic structures like stars and galaxies come from and how did they form?". The ceilings's immersive scenography displays a view into interstellar space featuring a variety of astrophysical objects, giving the visitors the feeling of being in and exploring space themselves.
Specific epochs in the Universe's evolution, which were governed by different physical phenomena, are represented by spatially separated sections via large walls (CMB  and inflation). Interactive exhibits let the visitors explore topics like star formation, dark matter, gravitational waves, astrophysical measurement techniques (gravitational lensing, optical telescopes) as well as the formation of baryonic matter (nucleons, atoms) from the hot and dense elementary particle soup in the early universe. The latter creates a clear link to the relevance of CERN's fundamental research.
\textbf{\emph{Exploring the Unknown}} follows a different approach. Here, open questions of modern physics such as the nature of dark matter and dark energy are addressed in the context of an art space which was designed together with CERN's arts programme \emph{Arts at CERN}. It displays artworks created by four internationally reknowned artists
who were inspired by the exhibition's overarching topics \emph{The Void}, \emph{The Invisible}, and \emph{Space \& Time}. These are complemented by audiovisual content in the form of interviews of CERN scientists answering questions about problems in modern physics, and of the artists talking about their motives and inspiration. The space gives the visitor the opportunity to experience science content outside the context of standard educational activities.
\vspace{-5pt}
\paragraph{Quantum World} In this space visitors are empowered to make sense of the puzzling principles of quantum mechanics through the help of interactive exhibits. These are centred around the topics of \emph{superposition}, \emph{interference}, \emph{entanglement}, \emph{tunnelling}, and \emph{the uncertainty principle}. They are featured as realisations of classic physics experiments like Young's double slit, or as fun and gamified  reinterpretations of these concepts.
Examples are the \emph{Quantum Tennis} exhibit where players need to hit a ball with uncertain velocity or position, or the \emph{Quantum Tunnelling} exhibit in which air hockey is turned into a challenge to shoot as many particles as possible through a physical barrier.
A key message for the visitor is that quantum mechanics cannot only be experienced in sophisticated lab experiments, but that it manifests itself in our everyday lives in many different ways. The unique part of this exhibition is the immersive space: it is an audiovisual experience in which groups of visitors equipped with headphones are guided through a path and advised to accomplish tasks together. This story-telling approach makes it possible to reiterate all the previously mentioned key principles in a coherent way.
\section{Key aspects of the exhibition development}\label{sec:accessibility_in_exhibition_development}
\begin{figure}[t]
      \includegraphics[width=\textwidth]{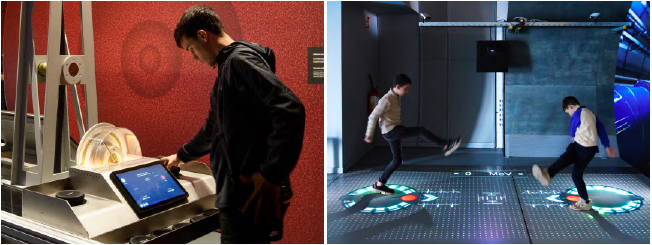}
      \caption{An example of two exhibits exploring the common question "What are nucleons made of?" at two different levels of complexity. \copyright CERN\\
      \emph{Left:} The more complex \emph{Proton Puzzle} exhibit featured in \emph{Back to the Big Bang}.\\
      \emph{Right:} The playful and social \emph{Proton Football} exhibit at the entrance of \emph{Discover CERN: Accelerate}.}
      \label{fig:pf_vs_pp}
\end{figure}

\begin{figure}[t]
  \includegraphics[width=\textwidth]{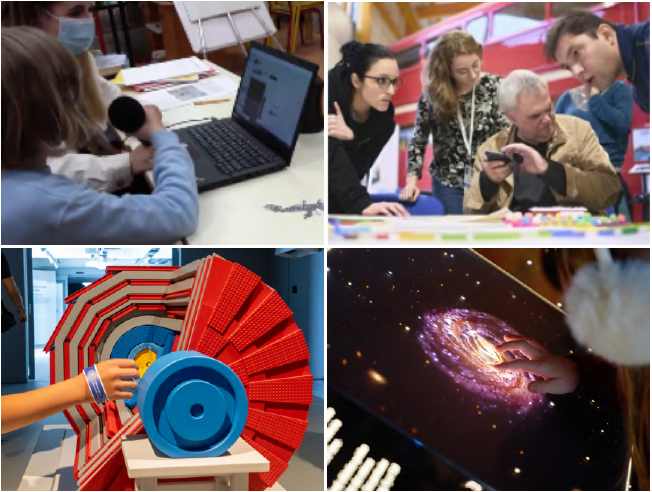}
  \caption{
    Collaborative prototyping workshops with different end-user groups in the content development phases of CERN exhibitions, and examples of accessibility improved exhibits in \emph{Science Gateway}. \copyright CERN\\
    \emph{Upper left:} Workshop with primary school children for the \emph{Science Gateway} exhibitions during the Covid pandemic.
    \emph{Upper right:} Workshop together with the \emph{Association for the Welfare of the Blind and Visually Impaired} (ABA) for the \emph{Microcosm} exhibition in 2018. 
    \emph{Lower left:} Different surface textures are used to represent the components of the detector model exhibit at the entrance of \emph{Discover CERN: Collide}.
    \emph{Lower right:} A tactile timeline guiding visitors through the history of our Universe in \emph{Back to the Big Bang}.
        }
    \label{fig:accessibility_tactile_elements_and_workshop}
\end{figure}

\noindent
The conception of the \emph{Science Gateway} exhibitions was based on more than 30 years of profound experience in exhibition development. 
It was shaped by the following overarching goals:
\vspace{-5pt}
\begin{itemize}[noitemsep]
  \item Get a broad and diverse audience to engage with the science and people working at CERN;
		\item Inspire the next generation to explore a career in science and technology;
		\item Empower visitors of all ages to make sense of the science that shapes their lives via a hands-on approach.
\end{itemize}
\vspace{-5pt}
Ensuring the broadest target audience possible requires taking into account various accessibility aspects during the content development phase.
This involves accounting for differences between visitor groups such as age, varying (scientific) educational backgrounds, nationality, socio-economic status as well as disabilities (blind, visual and hearing impairments, mobility difficulties).
We elaborate their implications in greater detail below.
\vspace{-10pt}
\paragraph{Age} The exhibition content is designed for visitors aged 8 years and above, starting from the age of reading capability.
Targeting such a young audience fosters early engagement with and interest in science and technologies, and creates memorable and fun experiences. This could later positively impact their choice for a career in STEM. The interactive and self-learning experiences at \emph{Science Gateway} playfully introduce visitors to the more abstract concepts of the scientific method and critical thinking.
\vspace{-5pt}
\paragraph{Varying educational background} Visitors may bring along very different levels of formal or informal (scientific) education when they enter the exhibitions. To tackle these different "starting positions" and facilitate engagement with the content, the key messages or topics of the exhibitions are often featured in multiple exhibits with ranging levels of complexity. One such example can be made by comparing the \emph{Proton Football} and the \emph{Proton Puzzle} exhibit (see Fig.~\ref{fig:pf_vs_pp}); both of them are linked to the question of the proton's inner structure.
At the first one, visitors are encouraged to first playfully collide protons then observe the result, whereas at the second one, one is challenged to find out the correct proton composition by comparing the constituents' masses on a scale. \emph{Proton Football} is a fun and social game for which no prior knowledge is needed to understand the exhibit, while at \emph{Proton Puzzle}, a certain degree of preknowledge about elementary particles (quarks, gluons) is required. 
\paragraph{Nationality} Ever since its foundation in 1954, CERN has been a truly international organisation welcoming employees and visitors from all over the world. Offering multilingual content in the exhibition is key to expanding the target audience to include international visitors, and make it accessible and possible to share with other CERN member states. The contents are available in five different languages (French, English, German, Italian, Spanish).
\vspace{-5pt}
\paragraph{Socio-economic status} The entrance to \emph{Science Gateway} and the participation in any of its activities is free of charge, making it easier for individuals and families with a lower income to explore and enjoy the research and outreach activities at CERN.
\vspace{-5pt}
\paragraph{Special needs due to disabilities} The representation of abstract concepts traditionally relies on 2D visualisation. Naturally, this leads to a lack of valuable experiences for the blind and visually impaired. 
An important step towards more accessible exhibition content is \emph{including} different end-user groups in the (early) content development stages such as by hosting a series of collaborative ideation and prototyping workshops (see Fig.~\ref{fig:accessibility_tactile_elements_and_workshop}). 
One result of this inclusive design process is the use of tactile elements on exhibits (see Fig.~\ref{fig:accessibility_tactile_elements_and_workshop}), either as part of the content itself, to highlight or distinguish important features, or to provide further information via tactile QR codes for corresponding audio content access, for example.
Furthermore, all video exhibits feature subtitles, which helps hearing-impaired visitors to engage with the content. 
Lastly, infrastructural improvements such as elevators, adjusted table heights, and well-distributed seating areas create an accessible and enjoyable experience for visitors with mobility difficulties.
\section{What do our visitors think? --- Statistics and preliminary evaluation results}\label{sec:stats_and_eval}
\subsection{Who are our visitors?}\label{sec:SG_statistical_overview}\noindent
CERN's \emph{Visits Service} continuously records data about the visitors of the various outreach activities. This allows deeper insights into the visitor group with respect to parameters such as age, educational background and the reasons for or the contexts of their visit. Here, we show the data collected from the example reference period of 1 June 2024 to 31 May 2025, taken from the CERN Public Outreach Database. During this time interval, \emph{Science Gateway} welcomed 400 838 visitors in total
of which 77\% visited the exhibitions. The additional information about this cohort is categorised as follows:
\vspace{-5pt}
\paragraph{"Types" of visitors} Most visitors either visited with their families, friends or as individuals (80\%), or came through an organised group visit (e.g. school or univeristy classes, teacher-student programmes at CERN) (19\%); 1\% of the visitors came in other configurations.
\vspace{-5pt}
\paragraph{Visit reasons/interests in the "Family \& individuals" category} The interests ranged from science (80\%) and technology (10\%) to CERN as an organisation (5\%) and international collaborations (1\%). Some visitors were simply in the area and looking for a fun activity (4\%).
\vspace{-5pt}
\paragraph{Visit reasons/interests in the "Group visits" category} The group visits largely took place in an educational (58\% school, 8\% university) and professional (13\%) context, but also as part of local tourism activities (14\%). The remaining 7\% named \emph{other} reasons.\label{statement:interest_in_science}
\subsection{Recent internal exhibition evaluation --- Visitors perception of exhibition accessibility}\label{sec:evaluation}
\begin{figure}[t]
     \centering
         \includegraphics[width=0.9\textwidth]{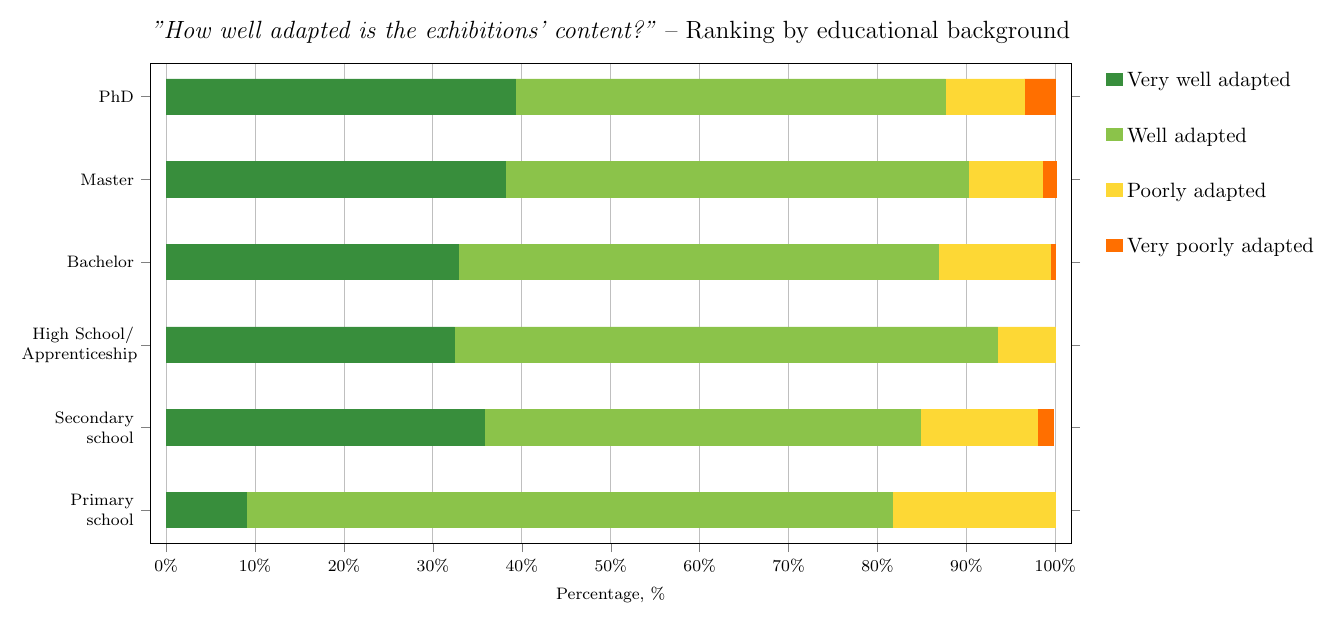}
        \caption{Survey results of the participants' perceived accessibility of the exhibition content, classified by their general educational background.}
        \label{fig:survey_accessibility_by_education}
\end{figure}

\paragraph{Set-up of the survey} To ensure the long-term quality and impact of the visitors' experiences, a team of internal staff has conducted a survey in the exhibition spaces from 30 July to 15 August 2024. The goal of the study was to obtain demographic information about the visitors (e.g. age, gender, educational background) and to learn about various aspects concerning their overall visiting experience (e.g. emotions/impressions connected to the exhibitions). The study followed a mixed-methods approach, combining qualitative methods (e.g. observing and tracking of visitors in the exhibition space) with quantitative (e.g. questionnaires) methods. The data was recorded at the exit of the exhibition, targeting visitors who have spent some time with the content and were leaving the space.
One caveat to the interpretation of this dataset is the chosen time frame of data collection. The summer months mean that the audience will be dominated by family visits and fewer organised school or other groups as usual. More precisely, about 90\% of the participants were part of a visit in a family setting. To counteract this issue and maintain representativeness, the authors of the survey evaluation have cross-referenced their data with a richer data set from CERN's \emph{Visits Service}. Note as well that the evaluation is not completed yet since the analysis process of the whole data set is only in its final stage.
\paragraph{Preliminary results --- How accessible are the \emph{Science Gateway} exhibitions?} The obtained data allows for a rich analysis of the overall visitors' experiences. However, here, we want to focus on the visitors' perceived accessibility of the exhibition content, and analyse how it might correlate with their educational background (see Fig.~\ref{fig:survey_accessibility_by_education}). As part of the questionnaire, the participants were asked the question "How well adapted would you rate the exhibition content on a scale from \emph{Very poorly} to \emph{Very well adapted}?". Here, a set of 807 responses in total is available. The corresponding overall result is relatively positive, with 54\% finding the content \emph{very well}, 35.2\% \emph{well}, 9.7\% \emph{poorly}, and 1.1\% \emph{very poorly adapted}. 
A subset of 789 responses can be further analysed with respect to the individual participant's general educational level.\newline
Regarding the distribution of the educational backgrounds within this subset, the survey data shows that 11.3\% of the participants have a PhD degree, 34.9\% have a Master's and 28.1\% a Bachelor's degree, 17.6\% attend/attended High School or do/have done an apprenticeship, 6.7\% are in secondary and 1.4\% in primary school. Thus, 74.3\% of all participants have an academic degree. Interestingly, the percentage of a negative response (\emph{Very poorly adapted}) seems to increase with increasing level of education, peaking in the PhD category with 3.5\%.
However, we observe that the majority of the participants (>80\%) in \emph{all} categories, with or without academic background, have found the contents to be \emph{very well} to \emph{well adapted}. It is clear that a higher educational background would facilitate the understanding of complex topics; therefore, the result in these particular categories is not very much surprising. It becomes even more clear when considering the large percentage of visitors coming (in family settings) with pre-existing interest in (and possibly knowledge about) science (see Sec.~\ref{statement:interest_in_science}). However, the largely positive result in the \emph{Schools and Apprenticeship} categories could indicate that the exhibition content has also been successful in targeting younger visitors as the exhibition developers aimed to. The content seemed to be best adapted for visitors who are/were visiting High School and/or do/have done an apprenticeship.
The percentage of responses judging the exhibition content to be \emph{very well} to \emph{well adapted} peaks at this category with 93.5\%, and decreases with decreasing educational level, given by 84.8\% and 81.7\% in the secondary and primary school category, respectively. The latter category also exhibits the total minimum percentage of \emph{very well adapted} responses, indicating that the content is \emph{by comparison} least adapted to the level of primary school students.
\vspace{-5pt}
\section{Conclusion}\noindent
Coming back to the initial question --- "Are Science Exhibitions for Everyone?" --- one may now ask whether the CERN \emph{Science Gateway} exhibitions succeed in targeting the broadest audience possible with regards to the accessibility efforts in the creation process. Combining the long-term data from CERN's Public Outreach Database and the preliminary results of the recent visitor survey, we draw the following conclusions:\newline
Most visitors are part of an individual and/or family visit, and of organised group visits with either a pre-existing interest (and knowledge) in science and technology or in an educational/professional context, respectively. Over 80\% percent of all survey participants find the exhibition content \emph{very well} to \emph{well adapted}.
The contents' adaptedness seems to be best evaluated by visitors at High-School or the Apprenticeship level, and evaluated the least well by visitors at the primary school level. 
Visitors with the highest formal education (PhD) have criticised the content's adaptedness the most (3.5\%).
The \emph{actual} audience still seems to largely consist of highly-educated science enthusiasts, who are more likely to spend a family trip in a science museum with their children. Therefore, it is important to continue efforts towards an improved outreach strategy to support groups with less formal education and science capital.\newline
Unfortunately, the recent exhibitions survey did not provide any information on the accessibility perception of visitors with disabilities, prohibiting an estimate of the impact of the inclusivity features on a variety of exhibits. Nevertheless, it is crucial to establish the integration of accessibility aspects into the exhibition development process to create valuable exhibition content and experiences for everyone.

\end{document}